# How to display science since images have no mass

Joël Chevrier[1,*], Hong Z. Tan[2], Florence Marchi[1], Gail Jones[3]

1-Institut Néel, CNRS/UJF 25 rue des Martyrs BP 166 38042 Grenoble cedex 9 France,
2-Haptic Interface Research Laboratory, Purdue University, West Lafayette, IN, USA
3-Department of Science, Technology, Engineering, and Mathematics Education, North Carolina State University, Raleigh, N.C.
* Author for correspondence

Education, science, in fact the whole society, extensively use images. Between us and the world are the visual displays. Screens, small and large, individual or not, are everywhere. Images are increasingly the 2D substrate of our virtual interaction with reality. However images will never support a complete representation of the reality. Three-dimensional representations will not change that. Images are primarily a spatial representation of our world dedicated to our sight. Key aspects such as energy and the associated forces are not spatially materialized. In classical physics, interaction description is based on Newton equations with trajectory and force as the dual central concepts. Images can in real time show all aspects of trajectories but not the associated dynamical aspects described by forces and energies. Contrary to the real world, the world of images opposes no constrain, nor resistance to our actions. Only the physical quantities, that do not contain mass in their dimension can be satisfactory represented by images. Often symbols such as arrows are introduced to visualize the force vectors.

### Forces are invisible

It is only with the use of combined visual and force displays that include a screen and a force feedback system or haptic device, that individuals can perceive in real time a virtual but consistent exploration of the represented reality. A haptic device can be defined, for this non-technical discussion, as a pointing device that responds to the displacement imposed by the user, by returning a force. For example, if the user moves a mass visualized on the screen, he or she will have to simultaneously apply the necessary real force to oppose the virtual weight.

Thanks to this technologically enhanced perception, anyone can explore inaccessible aspects of reality such as the nanoworld. But who cares about perceiving forces. Certainly science teachers who design nanoscale instruction care, as well as gamers, especially the ones focused on serious games and interactive interfaces. Beside in the field of education, one teaches teenagers that are connected to multimedia in average 7 hours a day as recently reported [1]. This means intense exposition essentially through screens to a truncated representation of reality. Going beyond, combination of visual and force displays can provide our perception with an immediate and reliable access to essential scientific ideas. We shall now examine how this applies to atom interactions.

The first lines of Galileo's book, the Starry Messenger [2] states: *«Great indeed are the things which in this brief treatise I propose for observation and consideration by all students of nature. I say great, because of the excellence of the subject itself, the entirely unexpected and novel character of these things, and finally because of the instrument by means of which they have been revealed to our senses.»* The instrument proposed by Galileo to all students of nature is of course the telescope. He insists on the fact that *«these things»* are revealed to our senses thanks to this instrument. For Galileo, the sense of interest is sight. Today, visual displays are everywhere and they carry images from space but also from the whole world. It is worth noticing here that our sense of hearing is also enhanced by advanced technologies that include loudspeakers and earphones. But touch is only now rapidly appearing with recent devices. The development of haptic tools, which can now add force displays to our perception, have increasingly been the focus of a technological research. Early haptic technologies resulted in an array of haptic devices with different degrees of effectiveness and widely different prices. Now some are highly advanced systems [3,4,5] and some are very inexpensive [6]. Although the less expensive devices are limited in their time response and the ability to represent smaller forces, they can still be used to introduce users to the existence of a force and to its key properties.

The question now arises: can force displays be part of a new, albeit virtual, Galileo-type of instrument? Are they opening a new frontier in the way we can perceive parts of reality that are normally remote to our perception?

**The goal of Galileo's telescope is for students of nature quite explicit:** *«In this way one may learn with all the certainty of sense evidence that the moon is not robed in a smooth and polished surface but is in fact rough and uneven, covered everywhere, just like the earth's surface, with huge prominences, deep valleys, and chasms.»*

### Hands on exploration of atomic interactions

Is there an equivalent of moon observation by the telescope that we can use for haptic displays? Where shall we put our hands (if technologically possible) so that the direct perception of a force will result immediately in new facts that we shall never forget, but that are usually taught only in advanced courses? This program has been written by Richard P. Feynman 50 years ago [7]: *"If, some cataclysm, all of scientific knowledge were to be destroyed and only one sentence passed on to next generations of creatures, what statement would contain the most information in the fewest words? I believe it is the atomic hypothesis (or the atomic fact, or whatever you wish to call it) that all things are made of atoms -- little particles that move around in perpetual motion, attracting each other when they are a little distance apart, but repelling upon being squeezed into one another."*

How can these points be turned into a shared and common knowledge? The first one seems fulfilled. Everybody is convinced that matter is made of atoms. This is certainly a great-success achieved in the past century. At school, one manipulates models of atoms and molecules with hands using the Balls and Sticks (B&S) model. There are numerous illustrations on the Internet of ways to represent perpetual thermal or Brownian motion. Not surprising, it is much more difficult to find ways to teach the associated Langevin random force. A challenge is to find ways to introduce trajectory and force characteristics of Brownian motion using combined visual and force displays. We shall not go any further in this direction here.

Ball and stick models represent many aspects of the atomic model such as atomic sizes, bond lengths and orientations but do not represent information about atomic interactions. A key limitation of ball and stick models is that they do not provide a direct perception of the attraction and repulsion between atoms, i.e. the energetic of the chemical bond.

At the human scale, we experience repulsion daily: objects are perceived as being separated and as not overlapping. This is essentially a consequence of the Pauli principle that indeed comes directly from the nature of the atom. Force displays coupled with Atomic Force Microscopes [8,9], either virtual or real, working in contact mode have been used over the last ten years to allow individuals to experience repulsion at different scales. This repulsion is essentially as equally forceful at all scales when transferred by force displays. This scale invariance does not change our way of looking at object interactions whatever their sizes. Indeed this is why B&S models can be representations of atom combinations and can be used so simply at our scale. It is primarily the sharpness of the atomic repulsion that allows us to model atoms as balls [10]. Atomic repulsion is implicitly represented in our fingers when we manipulate B&S models. Direct contact between objects at the human scale is essentially due to the same atomic repulsion at the nanoscale. When using B&S models, our perception of atoms is then not challenged.

To the contrary, attraction between objects that comes from atom interaction is not accessible to our senses without the help of an instrument. Attraction between atoms and nanoobjects is an essential consequence of atom behavior and is universal at the nanoscale. Students do not typically learn about this universal behavior of matter at the nanoscale until they reach an advanced level of study. The gecko lizard, now the star of the nano world, spectacularly manifests the importance of nanoscale attractive forces. This lizard, with its nanostructured feet can walk on the ceiling [11]. It shows at the human scale that there must be an attraction that we cannot perceive. New and detailed images of the nanostructured feet have led to the development of new nanotechnologies and associated adhesives.

### Force displays as educational tools [12,13,14]

Using the force displays with visual displays, a virtual haptic B&S educational module can allow for the hands-on perception of atomic interactions. Once used, one does not tend to forget that there is an attractive interaction between atoms and between nanoobjects.

We can now experience with haptic modeling the forces that exist as we move atoms or molecules closer to each other, first modeling the attractive van der Waals interaction then representing the much stronger iono-covalent attraction characteristics of chemical bonding and finally representing short range repulsion. If used in science museum for a large audience, people can explore atom interactions and the associated applications such as molecular bonding at the origin of SOI silicon wafers [15]. If embedded in science instruction, students can now use haptic computer technologies to have a direct and personal experience with atomic interactions as they build molecules on screen.

A machine that enables the user to experience non contact interaction at nanoscale [16,17] has already been built for the Expo Nano public exhibition in Europe [18]. The user manipulates either a virtual blind cane at our scale, the «macrostick», to touch a hard surface (no attraction involved) or a nanostick with a limited stiffness to touch a substrate at the nanoscale. The macrostick is a reference with no surprising behavior whereas the nanostick irreversibly jumps to contact as the user approaches the stick too close to the surface. The visitor experiences the types of attractive interactions that the gecko experiences when walking on the ceiling. Close to 200 000 of people have used this new tool across Europe. In other contexts, students in middle and high school have used haptic devices to manipulate and experiment with materials at the nanoscale through an Internet connection to an atomic force microscope[19].

This use of force displays combined with visual displays to manually explore acting forces in science is not limited to the nanoworld. It can be used to discover a new force and its properties (buoyancy, Lorentz force, quantum confinement or Brownian motion...). It is a new tool to help building the visual representation of a force as a vector [20].

The advancements with new haptic tools allow educators at all levels to open a new world of science to students and to ultimately help these students directly face the forces and their properties that lie beyond human perception although they are central in science and technology.

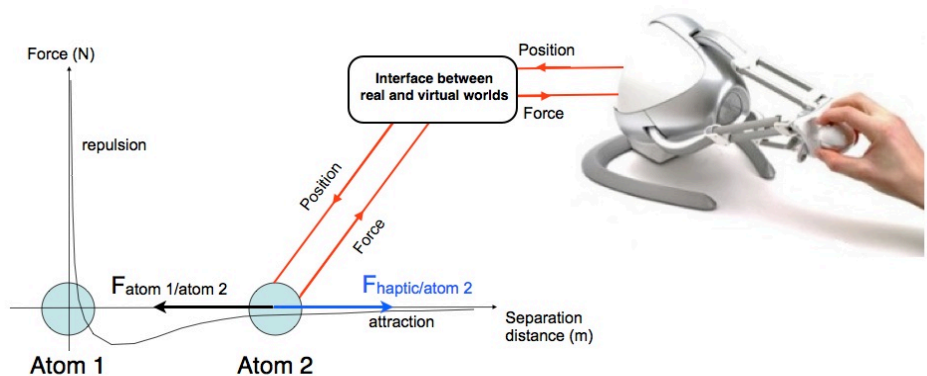